# Shear Flow in Cylindrical Open Channel Under Precession

Hajar Alshoufi[*1]

[1] Department of Hydraulic and Water Resources Engineering, Faculty of Civil Engineering, Budapest University of Technology and Economics, Műegyetem rkp. 3. 1111, Budapest, Hungary
[*] Corresponding author, e-mail: hajar.alshoufi@emk.bme.hu

**Abstract**
The study of forced oscillations in open cylindrical channel under precession is extended to include the shear effect, that is induced by inertial waves in such systems. The linear part of the problem led to two equations for stability one for the viscous part similar to Orr-Sommerfeld equation and one for the inviscid part similar to Rayleigh equation, the second was solved and discussed depending on the stream function observation. The linear part also led to relationship that connects between the stream velocity and the disturbance one is derived in a form similar to Burns conditions for open flows under normal conditions. Experimentally measuring the horizontal velocity distribution with depth showed that this distribution is sinusoidal one. Burns condition was solved based on this assumption. The nonlinear part of the problem led to a new version of Koteweg De-Vries (KdV) equation that is solved numerically by applying the leapfrog method for time discretization, Fourier transformation for the space one, and the trapezoidal rule for solving the integrals with depth, the results showed that the shear has no specific impact on the wave form which is similar to the classical results obtained by the theories under normal conditions.

**Keywords**
precession, open channel flow, shear flow, Rayleigh stability, KdV equation

## 1 Introduction

Most of the waves theories focus on the cases where the flow is irrotational one, from which the effect of stream vorticity is negligible. However, in real streams this is not the case, the shear effect induced by the friction effect due to water motion along the sides and the channel bottom modifies the whole stream velocity distribution with depth and makes it rotational one. Generally, due to the rotational action a thin layer called the "critical layer" may take place, this layer has thickness of order the amplitude of the wave [1], that is propagating along the stream, such layer only occurs when the disturbance velocity equals to the stream's one, that are connected with each other in the form of the famous condition introduced by Burns in [2]:

$$\frac{1}{g} = \int_0^z \frac{dz}{\left(U(z)-c\right)^2}, \qquad (1)$$

where $U(z)$ the stream velocity distribution as a function of depth $Z$, $c$ the wave velocity. The direct integration of Eq. (1) leads to the fact that wave velocity relative to the bottom of the stream has two values, of which one is always less that the minimum stream velocity (i.e., the velocity of the slip $U_{\min}$), and the other is always greater than the maximum stream velocity (i.e., the velocity at the surface $U_{\max}$), from which one can find directly that if the slip velocity is zero, then the wave velocity relative to the bottom has a negative value, so that is always upstream propagation of waves, which contradicts the shooting flow theory, in which all disturbances are swept downstream [2]. The variation of wave speed was noticed earlier in [3] by Thompson when the current is not uniform horizontally, the wave speed varies accordingly and distorts the wave fronts in a refraction pattern, thus the topography near the shore (which is considered shallow) is highly important on wave form as the shallow currents are essentially rotational ones, other important result he extracted that the phase speed velocity of waves in a current whose shear is constant differs from that of irrotational wave depending on the magnitude of shear. Busse [4] stated that the change in depth or any slight modification of uniform flows causes stretching of vortex lines similar to the effect of varying Coriolis force, the author discussed theoretically the case when small disturbance deviated from the solid-body rotation on shear flow, where a critical value of Rossby number (the definition can be found in [5]) over which the flow becomes unstable where the inertial forces overcome the friction in Ekman boundary layers and the constraints imposed by the variation in depth of the container. The stream velocity variation with depth is varied, Hunt [6] for instance assumed that the steady stream velocity to vary as the one-seventh power of the relative height above the bed. Johnson [7] presumed several shear functions including linear, Poiseuille, and Cubic parabolic profiles with discontinuous vorticity $U'$, where he solved Eq. (1) accordingly through the direct integration, with discussion of critical layer formation and the corresponding critical depth. Yih [8] discussed the different results of Burns condition and proved the case where the disturbance velocity is smaller than the slip one or the velocity at the bottom, the cases he discussed depended on the derivatives of the assumed stream velocity function, for instance he stated that it is impossible to get singular neutral mode from which $c = U$, where $c$ the disturbance velocity, if and only if the second derivative of $U$ does not change sign and $U$ is monotonic in the field of flow. Drazin and Howard [9] presented as well many schemes for the stream velocity distribution including the sinusoidal form, which is similar to the case in the present flume under study, also they discussed the rectangular jet form, plane Couette flow. Their starting point was the modified version of Navier-Stokes equations that takes the perturbation effects, from which they discussed boundary layer formation for inertial waves, and different cases of instability and the corresponding solutions. Fenton [10] tried to discuss the problem in different manner from which he expanded the kinematic and dynamic conditions in terms of the stream function and solved the problem numerically from which a new dispersion relationship between the wave number and celerity was extracted in the dimensionless form in terms of Froude number that was assumed in terms of shear velocity. As the conclusion of Burns was the importance of viscous effect to be included, other team Velthuizen and Wijngaarden [11] figured out that the condition that looked physically unacceptable can be accepted if the wave velocity is complex one, which implies damping or growth of the waves, which may be caused by viscous effects. They tried to derive the condition of Burns in different way by connecting wave kinematic and potential energies using the linearization of Bernoulli equation after stream flow velocity inclusion. Johnson [1] derived the main equations in three regions inside the critical layer (which occurs when $U = c$), above and below it. He assumed initial configuration to contain no closed streamlines so that the vorticity can presumably be assigned from the undisturbed conditions at infinity. He justified the nonlinear approach inside the layer if the amplitude parameter is greatly in excess of the inverse Reynolds number. Assuming irrotational effects to study the solitary wave without invoking the shear flow looks as a shortcoming that was first amended by Brooke Benjamin [12] who was the first in deducing the solitary wave over long channel of a nonuniform parallel stream when vorticity produced by frictional action at the boundary becomes diffused over the whole cross section. The author suggested that instead of zero vorticity everywhere (irrotational flow), there is constant vorticity along each streamline, provided the limitations due to stagnation effects. Later both Freeman and Johnson [13] also provided the derivation of solitary wave and its solution taking the shear flow into consideration. The aim of this work is to extend the previous linear and nonlinear work that was carried out by the present author to add the shear effect. To add the proper distribution of velocity with depth measurements were carried out using Vectrino Profiler (ADV) at specific point in the channel with depth, it turned out that the form vertically is sinusoidal one. This function was inserted into the new version of Burns condition that was derived for the linear case and solved accordingly, similar to the normal flows the wave velocity relative to the bottom has two values negative and positive. The nonlinear part of the problem aims first on deriving new KdV model that includes the shear effect and then to figure out what is the possible effect of the shear on the wave form, is it like the normal cases derived in [12] where the potential flow still valid? Or does the shear here force us using different formulae? It turned out that no impact appears.

The paper is divided as follows in the §2 a deep explanation about the experimental setup and the governing equations, in §3 a dedicated discussion on the linear periodic case and the stability using Rayleigh equation is elucidated. In §4 the nonlinear part of the problem and the new KdV model is derived. In §5 discussion on the numerical scheme using Fourier transformation and the leapfrog methods. In §6 the main conclusion and summarization of the whole results.

## 2 Experiment setup
The structure of the apparatus is shown in Fig. 1. The walls of the circular wave flume are two coaxial cylinders both made of Plexiglas. They are based on a round shaped plastic bottom plate, which is fixed to a rigid wooden round support table of the same radius (60 cm). We refer this as the *tilting table* because the center of this table is mounted on a support column through a Cardano type universal ball joint. The joint contains two, mutually perpendicular horizontal axles, around which the table can freely tilt in any direction, but it prevents any rotation around the vertical axis. This type of joint is used to transmit torque in between two nonaligned shafts in different machineries (like automobiles). Right below the tilting table there is another horizontally positioned round table of the same size, which is capable of free rotation around the vertical axis. The tilting table is partially supported also by the rotating table on three points arranged in 120° apart from each other. Each of this support consist of a vertical adjustable length spreader screw mounted vertically on the rotating table and a roller on top of the screw, on which the tilting table rests. By carefully adjusting the spreader screws, the tilt angle and the direction of the tilt can be set. When the lower table turns, the tilt direction in which upper table - together with the water flume - tilts also turns around (the vertical axis) without changing its slope or turning around (the vertical axis). This motion is called *precession* in classical mechanics and is well known, e.g., in celestial mechanics. The final bit of the apparatus is an adjustable speed direct current (DC) electric motor, which drives the rotating table via a belt transmission system and a vertical cylinder shaft turning on two bearings coaxially around the main support column. We note that currently the rotating table can be turned only in the counterclockwise sense. The main sizes and the crucial parameters of the system are summarized in Table 1.

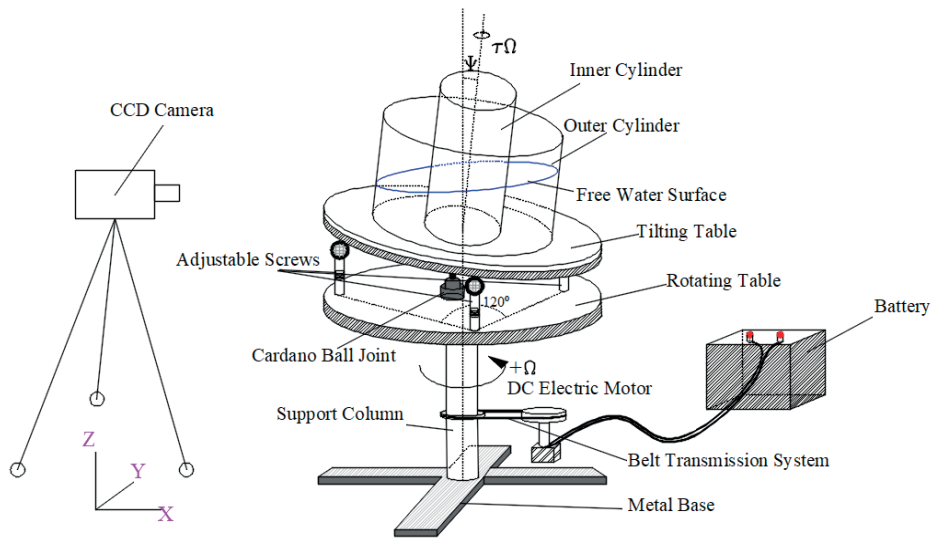

**Fig. 1** Sketch of experimental setup. The camera system in front of the flume almost at the same level of the water level inside the tank

The cylinders rotate about their common axis $(O, \hat{k})$ with precession rate $\Omega_2 = \tau\Omega$, where $\tau$ the slope of the base table. The lower table is able to rotate about an axis $(O, z)$ with rotation rate $\Omega$. The angle between the local axis $\hat{k}$ and the inertial one $z$ is the angle of precession $\psi$. The cylinders are filled with water of different volumes, the inner cylinder has higher height than the outer one. The cylindrical coordinates suit the geometry of the flume but with proper conversions between the inertial laboratory floor coordinates to the tilting table should be taken into consideration so that the final system of Navier-Stokes equations is given as:

$$\frac{d\vec{\tilde{V}}}{dt} + \vec{\tilde{V}}\cdot(\nabla\vec{\tilde{V}}) = G - \frac{\nabla P}{\rho} - 2\tilde{\Omega}(\Omega t) \times V'' + \nu\nabla^2\tilde{V}, \qquad (2)$$

where $\tilde{V}$ the velocity vector, $\tilde{\Omega}$ the angular velocity vector, the tilde over the variables indicates to their values in the tilting rotating frame, $P$ the pressure, $t$ the time, $G$ the gravity force, $\rho$ the density, $\nu$ the kinematic viscosity. With continuity equation:

$$\nabla \tilde{V} = 0 . \tag{3}$$

**Table 1** Geometrical information of the channel

| Notation | Value | Description |
| --- | --- | --- |
| $R = r_{max}$ | 223 mm | Outer Radius |
| $\beta R = r_{min}$ | 125 mm | Inner Radius |
| $b = r_{max} - r_{min}$ | 98 mm | Channel Width |
| $\beta = r_{min}/r_{max}$ | 0.5605 | Radial Ratio |
| $h$ | 260 mm | Maximum Wall Height |
| $A$ | 10.71 dm$^2$ | Base Area |
| $\forall$ | 27.5 dm$^3$ | Maximum Volume |
| $\tau$ | 0.005… 0.12 | Tilt |
| $\bar{h}$ | 20… 140 mm | Mean Water Level |
| $\Omega$ | 1.5… 8 rad/s | Angular Velocity |
| $z_o$ | ≈ 50 mm | Elevation of Flume Bottom |

The impermeability boundary conditions on the outer and inner radii, and the bottom:

$$\tilde{V} = 0 . \tag{4}$$

The motion inside the flume is assumed to be two-dimensional with predominant direction the azimuthal one, where Katsis and Akylas [14] stated that in a channel of finite width $b_1 = \frac{b}{2}$ three dimensional effects are negligible if $b_1 \leq \bar{h}^2/a$, where $\bar{h}$ the average depth of water, $a$ the typical wave amplitude. Table 2 shows results from the experiment about this ratio, where the experiments focused on relatively big amount of water inside, those are the cases where the solitary wave was noticed, as the final aim in this work is to compare the extracted KdV equation under shear effect with the experiment. The tilt angle characterized by the slope $\tau$ also was small. It was noticed during the whole experiments that the shallowness condition is satisfied, or we can assume that the wave motion is propagating into a shallow water as in Table 2, where the shallowness parameter is given as: $\delta = \left(\frac{H}{R}\right)^2 \ll 1$. where $H = \bar{h}$ represents the depth scale, $R = r_{max}$ represents the length scale or the horizontal scale. The vertical range of the flow takes place from the bottom where $z_0 = 0$, till the free surface where when no disturbance applied to the surface is assumed to be the average depth of water $z = \bar{h}$, by which we satisfy the mean water level constraint:

$$V = \int_0^{2\pi} \int_{r_{min}}^{r_{max}} \int_0^{\bar{h}} r dz.dr.d\theta = \bar{h}\left(r_{max}^2 - r_{min}^2\right)\pi . \tag{5}$$

### 3 Periodic linear case

The mean velocity is assumed in the main flow direction and a function of water depth, so that the final linearized version of Navier-Stokes equations: (6)

$$u_t + \frac{Uu_\theta}{r} + wU_z = g\tau \sin(\Omega t - \theta) - \frac{P_\theta}{\rho r} - 2\Omega\tau\cos(\Omega t - \theta)w$$
$$+\vartheta\left(\frac{u_{\theta\theta}}{r^2} - \frac{u}{r^2} + u_{zz}\right),$$
(7)

$$w_t + \frac{Uw_\theta}{r} = -\frac{P_z}{\rho} + 2\Omega\tau\cos(\Omega t - \theta)[u+U] + \vartheta\left(\frac{w_{\theta\theta}}{r^2} + w_{zz}\right),$$

where $u, w$ the azimuthal and axial velocities, respectively. $P$ the pressure. With mass conservation:

$$u_\theta + rw_z = 0 . \tag{8}$$

**Table 2** Two-dimensional and shallowness effects in the channel

| $V$(ml) | $a$(m) | $\bar{h}$ (m) | $b_1$ (m) | $\bar{h}^2/a$ (m) | $\tau$ | $\delta$ |
|---|---|---|---|---|---|---|
| 10000 | 0.112 | 0.0933 | 0.049 | 0.078 | 0.0167 | 0.175 |
| 12000 | 0.0886 | 0.112 | 0.049 | 0.142 | 0.0117 | 0.252 |
| 14000 | 0.1024 | 0.131 | 0.049 | 0.167 | 0.0233 | 0.344 |

The boundary conditions on both the bottom and the free surface can be written as:

$$w = \eta_t + \frac{U\eta_\theta}{r}; \quad w = 0, \tag{9}$$

where $\eta(\theta,t)$ the free surface function. $U$ the stream velocity. Other condition related to the assumption of steady constant vorticity at specific streamline, thus as we assumed two-dimensional motion the stream functions are given as: $u = \varphi_z, w = -\frac{1}{r}\varphi_\theta$. So that the vorticity dynamical condition is:

$$\frac{1}{r^2}\varphi_{\theta\theta} + \varphi_{zz} = -\xi , \tag{10}$$

where $\xi = \left(\frac{1}{r}\frac{\partial w}{\partial \theta} - \frac{\partial u}{\partial z}\right)$ the constant vorticity. By cross-differentiating the Eq. (6) and Eq. (7) pressure may be eliminated to reduce the system into final equation: (11)

$$\frac{\partial}{\partial t}\left(u_z - \frac{w_\theta}{r}\right) + U''w + \frac{U}{r}\frac{\partial}{\partial \theta}\left(u_z - \frac{w_\theta}{r}\right) + 2\Omega\tau\sin(\Omega t - \theta)\frac{u}{r}$$
$$+\vartheta\left(\frac{w_{\theta\theta\theta}}{r^3} + \frac{w_{zz\theta}}{r} + \frac{u_z}{r^2} - u_{zzz} - \frac{u_{\theta\theta z}}{r^2}\right) = 0.$$

Equation (11) shows an additional sinusoidal term that appears because of the axial Coriolis force, as the azimuthal one was cancelled in the derivation, the number of primes over the unknown quantities indicates to the order of derivatives in the vertical direction of the flow $\frac{\partial}{\partial z}$. Let's assume some stream function of Fourier mode that has azimuthal decay as follows: $\varphi(\theta, z, t) = \phi(z)e^{ik(\theta - \Omega t)}$, where $k$ the wavenumber, $\Omega$ the solid-body rotation velocity projected at the outer radius of the channel from which any disturbance is assumed to move with it. Substituting into Eq. (11) will give:

$$W\phi'' + \frac{2\Omega\tau i}{k}\sin(\theta - \Omega t)\phi' - \left[\frac{k^2}{r^2}W + U''\right]\phi =$$
$$\frac{i r \vartheta}{k}\left[\phi'''' + \frac{k^4}{r^4}\phi - (2k^2 + 1)\frac{\phi''}{r^2}\right],$$
(12)

where $W = U - r\Omega$. This equation is the forced Orr-Sommerfeld equation under precession conditions. Equation (12) is fundamental for stability of laminar flows in cylindrical channel under precession. When assuming inviscid conditions like the case under study where water used in the experiment, the right-hand side of this equation can be neglected, the equation can be written:

$$W\phi'' + \frac{2\Omega\tau i}{k}\sin(\theta - \Omega t)\phi' - \left[\frac{k^2}{r^2}W + U''\right]\phi = 0 \tag{13}$$

Equation (13) is called the Rayleigh stability equation that takes the precession effect in the closed circular flume. This equation has variable coefficients in $z$, $\theta$, $t$.

### 3.1 Stream function observations

The measurements of velocity stream profile with depth was done by the help of Acoustic Doppler Velocimeter (ADV), the probe has four receiver arms and central one called the transmit transducer, during the experiment it was approximately in the middle of the channel where it was hold on a metal rod that is supported on other metal procedure that is fixed to the laboratory floor, the Vectrino while measuring does not feel the effects of rotation or tilt. In order to avoid the bad bed effects it is better to rise the Vectrino from the channel bed about three centimeters, thus the lowest depth where the measurements started was about 3 cm. The Vectrino also should stay far from the free surface where the air effect makes the signal chaotic and not true, thus also we left about 1–2 cm from the top, so that the effective depth is about 5–7 cm, but this also depends on the average amount of water in the channel, as we tried to measure for relatively big amounts of water to tackle the problems from the bottom and the surface. Sediment material like sand were also seeded so that we enhance the echo. (Details of Vectrino measurements can be found in a previous work by the present author [15]) The cases were tracked are three: the first is when the free surface just simple closed circles that rotate and tilt with slow energy provided to the system the profile has sinusoidal form as it is clear in Fig. 2(a), the second case is when the single Kelvin solitary wave appears with high power provided to the system (resonance conditions) as it is clear in Fig. 2(b), it even has more or less Gaussian function form, and finally the case of simple sinusoidal waves appear (three of them) as in Fig. 2(c). In the first case the wave motion is slow thus the horizontal velocity goes to and fro many times crossing the vertical line with relatively equal amplitudes in the positive and negative directions of motion with many inflection points, on the contrary for the second case where the wave motion is faster the wave phase is small thus the amplitude is bigger in the positive direction of the flow with many smaller amplitudes in the negative direction and many inflections in the curve. The stability can be studied based

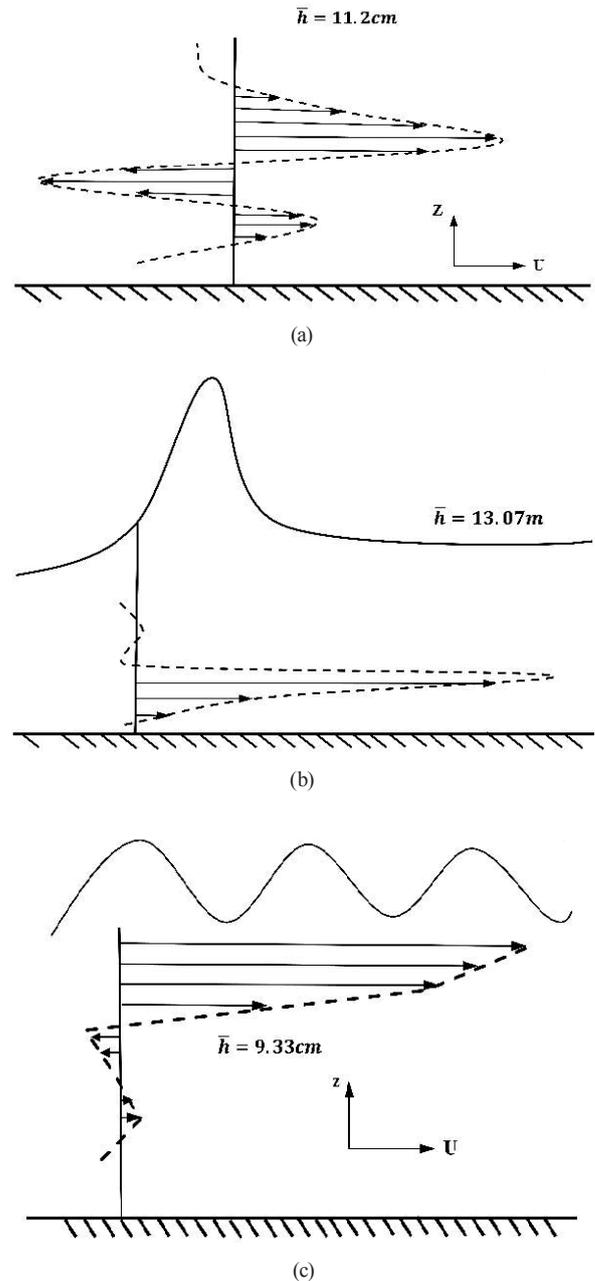

**Fig. 2** (a) Stream velocity distribution with depth in case no motion on the surface $\bar{h}$ = 11.2 cm, $\Omega$ = 1.15 rad/s, $\tau$ = 0.005; (b) in resonance case, $\bar{h}$ = 13.07 cm, $\Omega$ = 6.48 rad/s, $\tau$ = 0.01333; (c) case of sinusoidal waves on the surface, $\bar{h}$ = 9.333 cm, $\Omega$ = 4.18 rad/s on the stream function and of course Rayleigh equation, where the final solution will give us information if the flow is stable or not. The cases in Fig. 2 all can be covered by the following assumption of the stream function:

$$U(z) = U_1 \sin(mz), \tag{14}$$

where $m$ the axial wave number. $U_1$ the amplitude. By substituting the function into Forced Rayleigh stability

Eq. (13) we get:

$$(U_1 sin(mz) - r\Omega)\phi'' + M_1\phi' \\ - \left[\frac{k^2}{r^2}(U_1 sin(mz) - r\Omega) - m^2 U_1 sin(mz)\right]\phi = 0, \quad (15)$$

where $M_1 = \frac{2\Omega\tau i}{k}sin(\theta - \Omega t)$. To simplify the problem, we assume a balance between the phase of tilt and the azimuthal angle so that we get rid of the Coriolis terms, and we get a simple homogeneous equation:

$$(U_1 sin(mz) - r\Omega)\phi'' \\ - \left[\frac{k^2}{r^2}(U_1 sin(mz) - r\Omega) - m^2 U_1 sin(mz)\right]\phi = 0. \quad (16)$$

Second, we assume long wave approximation so that $k \to 0$, then the final simplified version takes the form:

$$(U_1 sin(mz) - r\Omega)\phi'' + m^2 U_1 sin(mz)\phi = 0. \quad (17)$$

The vertical depth is confined between the bottom $z = z_1 = 0$, and the free surface which is the average depth of water in case of no motion $z = z_2 = \bar{h}$, although the whole channel is wobbling, so the observer can also see the sediment particles that were added to enhance the echo signal moving to and fro at the bottom of the channel but their velocity is relatively small in comparison with bigger ones close to the free surface, thus if we assume zero bottom conditions and neglect $r\Omega$, the equation is reduced into:

$$\phi'' + m^2\phi = 0. \quad (18)$$

Which can be solved using the elementary methods to get the solution:

$$\phi = A_1 cos(mz) + A_2 sin(mz), \quad (19)$$

where $A_1$, $A_2$ some constants, that can be determined from the boundary conditions, at $z = 0$, $\phi(0) = 0$, thus $A_1 = 0$, at $z = \bar{h}$, $\phi(\bar{h}) = 2\pi r\Omega\bar{h}$ in case no waves on the surface thus the final solution:

$$\phi = \frac{2\pi r\Omega\bar{h}}{sin(m\bar{h})} sin(mz). \quad (20)$$

However, for the case where $k \neq 0$, and assuming that the depth velocity almost zero then Eq. (16) can be written as:

$$U_1 sin(mz)(D^2 - \alpha^2)\phi + m^2 U_1 sin(mz)\phi = 0, \quad (21)$$

where $\alpha^2 = \frac{k^2}{r^2}, D^2 = \frac{\partial^2}{\partial z^2}$. Following the way introduced in [6] we take:

$$D^2 U(z) = 0, \quad U_1 sin(mz) = 0, \quad (22)$$

$$z = z_s = \frac{n\pi}{m}, \quad (n = 0, \pm 1, \pm 2, ...), \quad (23)$$

where $z_s$ the depth accords with inflection point existence. Now depending on the number of inflection points the flow can be either stable or unstable and this depends on the length of the axial domain where the motion exists. The stability solution that satisfies Eq. (21) then is given:

$$\phi_s = sin\left(\frac{n\pi}{m}\frac{z}{\bar{h}}\right), \quad (24)$$

$$\alpha_s = \sqrt{m^2 - \frac{n^2\pi^2}{m^2\bar{h}^2}}, \quad (25)$$

where $\alpha_s = \frac{k}{r}$, which is for long waves $\alpha_s = 0$, then we can write $n = \frac{m^2\bar{h}}{\pi}$. If $m^2\bar{h} > \pi$ the flow is unstable, and stable if $m^2\bar{h} < \pi$. If we assume that $m = 1$, then only if $\bar{h} < \pi$ the flow is stable where no inflection point can appear in this case, but this contradicts the experimental results as even when taking small depth the inflection points appear, the reason is that the channel all the time wobbling thus, we cannot insure this. When assuming the whole vorticity Eq. (13) and assume fixed time and azimuthal angle, we find:

$$W\phi'' - [\alpha^2 W + U'']\phi + M\phi' = 0, \quad (26)$$

where $M = \frac{2\Omega\tau i}{k}sin(\theta - \Omega t) = const$. If we apply the equation on the free surface where $z = \bar{h}$, the coefficients will be all constants and taking Eq. (14). Then the equation is written:

$$A\phi'' + B\phi' + C\phi = 0 \quad (27)$$

$A = W$, $B = M$, $C = -[\alpha^2 W + U'']$, this is quadratic equation, the delta is given by:

$$\Delta = -\frac{4}{K^2}\tau^2\Omega^2 sin^2(\Omega t - \theta) + 4\alpha^2 W^2 + 4WU''. \quad (28)$$

If $\Delta > 0$, the solution at the free surface:

$$\phi = C_1 e^{D_1\bar{h}} + C_2 e^{D_2\bar{h}}. \quad (29)$$

If $\Delta < 0$, the solution is:

$$\phi = \left[C_1 cos(\beta\bar{h}) + C_2 sin(\beta\bar{h})\right]e^{\varpi\bar{h}}, \quad (30)$$

where the complex roots are: $\varpi \pm i\beta$. The stability condition in this case shows that if $\varpi < 0$, any disturbance or perturbation on the free surface will vanish with time, and it is stable, on the other hand if $\varpi > 0$ the solution will grow up with time, and it will be unstable [16].

## 3.2 Burns condition and its solution

The previous analysis was carried out on assuming that the disturbance velocities and their derivatives are all small quantities from which all nonlinear terms were neglected. By introducing the dimensionless variables to the system of Euler equations (assuming that the fluid is inviscid) this will factor out all Coriolis forces at the leading order of the problem as follows:

$$t \to \frac{\bar{t}}{\Omega}, \quad r \to R\bar{r}, \quad \eta \to \varepsilon H\bar{\eta}, \quad z \to H\bar{z},$$
$$u \to [\varepsilon R\Omega \bar{u} + U], \quad w \to \varepsilon H\Omega \bar{w}, \quad \tau \to \sqrt{\delta}\bar{\tau}, \quad (31)$$
$$P = P_a + \gamma(H-z) + \varepsilon \rho \Omega^2 r^2 \bar{P}, \quad \theta \to \bar{\theta}, \quad \Omega \to \bar{\Omega},$$

where $\varepsilon = \frac{A}{H}$, the amplitude parameter. On assuming steady case where the coordinates move with the wave, and as the final aim is to derive the KdV equation which is used to describe the waves of permanent form, and those waves undergo slow changes in form thus slow time evolution we introduce the following:

$$\frac{\partial}{\partial t} = -\Omega \frac{\partial}{\partial \bar{\theta}} + \varepsilon \frac{\partial}{\partial T}. \quad (32)$$

Then Euler equations will be written after introducing the shear effect and dropping the primes:

$$\frac{u_\theta}{r}(U - r\Omega) + wU' + \varepsilon\left(u_T + \frac{uu_\theta}{r} + wu_z\right) =$$
$$-\frac{P_\theta}{r} + \frac{\tau}{F_r}\sin(t-\theta) - \varepsilon[2\tau\cos(t-\theta)w], \quad (33)$$

$$\varepsilon\left[\frac{w_\theta}{r}(U - r\Omega) + \varepsilon\left(w_T + \frac{uw_\theta}{r} + ww_z\right)\right] =$$
$$-P_z + 2\tau\cos(t-\theta)U + \varepsilon[2\tau\cos(t-\theta)u], \quad (34)$$

$$u_\theta + rw_z = 0, \quad (35)$$

$$w = \varepsilon\left(\eta_T + u\frac{\eta_\theta}{r}\right) + (U - r\Omega)\frac{\eta_\theta}{r}. \quad (36)$$

$F_r$ is Froude number of the problem that is assumed: $F_r = \frac{H\Omega^2}{g}$. The prime in Eq. (33) indicates derivative in terms of the vertical direction $\frac{\partial}{\partial z}$. It is worth to mention up to this point that the tilt angles in this system are in general small ones, and that the cases where the solitary wave were noticed corresponded to very small tilt character factor: $\tau$, (cf. Table 1, and Table 2), thus it is assumed that if $\varepsilon \to 0$ (which is the solution we are interested in for small wave amplitude) then $\tau \to 0$, so that at the leading order of the problem the forced terms can be neglected:

$$\begin{cases} (U - r\Omega)\frac{u_{0\theta}}{r} + w_0 U_0' = -\frac{P_{0\theta}}{r} \\ -P_{0z} = 0, P_0 = \eta_0 \\ u_{0\theta} + rw_{0z} = 0 \\ w_0 = (U - r\Omega)\frac{\eta_{0\theta}}{r} \end{cases}. \quad (37)$$

The azimuthal equation after substituting the stream function will give:

$$\phi'(U - r\Omega) - U'\phi = -\bar{\eta}. \quad (38)$$

Which is similar to the one derived by in [2] in Eq. (25) in the dimensionless form. After substitution of the value of the axial velocity into the azimuthal momentum, we get a new version of Burns condition, that connects between the stream velocity with the solid-body rotation of the flow $c = r\Omega$. We have to mention that the speed of inertial single oscillations that occur in this system have close value to the one of long wave speed, this was proposed in [17], in terms of the amplitude parameter $\varepsilon$ by:

$$c = \sqrt{g\bar{h}}\left(1 + \frac{1}{2}\varepsilon\right), \quad (39)$$

$$\int_0^z \frac{dz}{(U - r\Omega)^2} = 1. \quad (40)$$

Equation (40) is the dimensionless Burns Condition in case of channel system under precession. It is well known that this integral equation for $c = r\Omega$, certainly admits two solutions, as suggested in [7], for $c$ if $U(z)$ satisfies $U'(z) > 0$, and $U''(z) < 0$: one solution gives $c < U(0)$ at the bottom, and the other $c > U(1)$ at the dimensionless free surface at the leading order. To solve Burns condition we substitute the sinusoidal form Eq. (14) into Eq. (40) we get:

$$\int_0^z \frac{dz}{(U_1 \sin(mz) - c)^2} = 1. \quad (41)$$

By assuming $mz = x$, we assume also from trigonometric case that: $x = 2.\arctan(t)$, $t = \tan\left(\frac{x}{2}\right)$. From which we can write:

$$\sin(x) = \frac{2t}{1+t^2}, \quad dx = \frac{2dt}{1+t^2}. \quad (42)$$

After some mathematical manipulation we get an equation:

$$\frac{A.\tan\left(\frac{m}{2}\right)}{B_1\left[\tan\left(\frac{m}{2}\right) - B_1\right]} + \frac{B.\tan\left(\frac{m}{2}\right)}{B_2\left[\tan\left(\frac{m}{2}\right) - B_2\right]} + \frac{m}{2} = 0, \quad (43)$$

where:

$$B_1 = \beta + \beta\sqrt{1 - \frac{1}{\beta^2}}, \quad B_2 = \beta - \beta\sqrt{1 - \frac{1}{\beta^2}},$$
$$A = \frac{B_1^2 - 1}{B_1^2 - B_2^2}, \quad B = \frac{1 - B_2^2}{B_1^2 - B_2^2}, \quad (44)$$

where $\beta = \frac{U_1}{c}$, also it was assumed that $m = 1$. Equation (43) has no solution when $\beta = 0, 1$, and has single solution if $\beta > 0$, or if $\beta < 0$. By giving several values for $U_1$ and $c$ Eq. (43) can be solved accordingly as it is clear in Fig. 3:

## 4 Periodic nonlinear case

The final system of Euler equations after scaling Eqs. (33) to Eq. (36) will give at the zeroth order:

$$\begin{cases} w_0 = WI_2 \frac{\eta_\theta}{r} \\ w_{0z} = \frac{\eta_\theta}{r}(WI_2)' \\ u_0 = -\eta_0 (WI_2)' \\ u_{0z} = -\eta_0 (WI_2)'' \end{cases} \quad (45)$$

where $W = U - r\Omega$, $I_2 = \int_0^1 \frac{dz}{(U - r\Omega)^2} = 1$.

At the first order of the problem, all nonlinear terms appear again, which are all in terms of the zeroth order. So, we can write:

$$\begin{cases} (U - r\Omega)\frac{u_{1\theta}}{r} + w_1 U_1' + u_{0T} + \frac{u_0 u_{0\theta}}{r} + w_0 u_{0z} = \\ \quad \frac{\tau \sin(t-\theta)}{F_r} - \frac{P_{1\theta}}{r} - 2\tau \cos(t-\theta) w_0 \\ (U - r\Omega)\frac{w_{0\theta}}{r} = -P_{1z} + 2\tau \cos(t-\theta)[u_0 + U_0] \\ u_{1\theta} + rw_{1z} = 0 \\ w_1 + \eta_0 w_{0z} = (U - r\Omega)\frac{\eta_{1\theta}}{r} + \eta_{0T} + \frac{u_0 \eta_{0\theta}}{r} \end{cases} \quad (46)$$

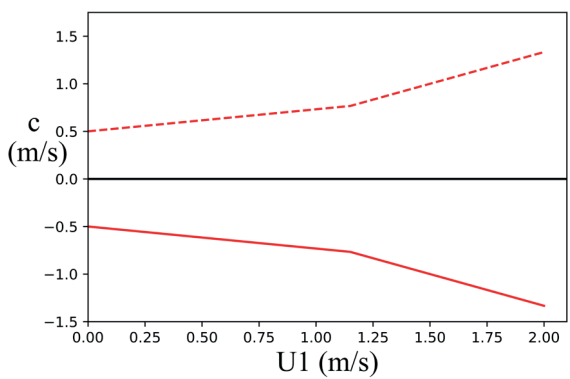

**Fig. 3** The solution for $c$ against surface speed $U_1$ for the sinusoidal profile, including $0 \le U_1 \le 2$

By applying the zeroth solutions into the azimuthal momentum and balancing the kinematic condition of the free surface with the azimuthal momentum. Then the final equation we get similar to the one extracted in [13] which suits the cylindrical geometry under study with additional forcing terms come from Coriolis and gravity forces:

$$\eta_{0T}\left[1 + W_1 \int_0^1 \frac{(WI_2)'}{W_1^2} dz\right] \quad (47)$$
$$+ \frac{\eta_0 \eta_{0\theta}}{r}\left[-2W' - \frac{2}{W} + W_1\left\{\int_0^1\left(\frac{WI_2(WI_2)''}{W_1^2} - \frac{(WI_2)'^2}{W_1^2}\right) dz\right\}\right]$$
$$+ W_1 \int_0^1 \frac{\tau \sin(t-\theta)}{F_r W_1^2} dz - W_1 \frac{\eta_{0\theta\theta\theta}}{r^3} \int_0^1 \int_z^1 \int_0^z \frac{W^2(z_2)}{W^2(z_1)W^2(z)} dz_2 dz_1 dz$$
$$+ \frac{2\tau W_1}{r}\sin(t-\theta)\int_0^1 \int_z^1 \left[\frac{U - \eta_0(WI_2)'}{W_1^2}\right] dz_1 dz$$
$$+ 2\tau W_1 \cos(t-\theta)\int_0^1 \int_1^z \left(\frac{(WI_2)'}{W_1^2}\frac{\eta_{0\theta}}{r}\right) dz_1 dz$$
$$- 2\tau W_1 \cos(t-\theta)\int_0^1 \frac{(WI_2)}{W_1^2}\frac{\eta_{0\theta}}{r} dz = 0$$

Which is the KdV version that takes the shear flow into consideration in an open cylindrical channel under precession. This equation will be solved numerically and compared with the experimental results. Before applying the stream function into the equations, we can do some proper integrations we find finally that the equation will take the form:

$$\eta_{0T}\left[2 + \int_0^1 \frac{W'}{W_1} dz\right] + \frac{\eta_0 \eta_{0\theta}}{r}\left[-W' - \frac{1}{W} - \int_0^1 \left(\frac{\frac{W'^2}{W} + 2W'}{W^2} + \frac{1}{W^3}\right) dz\right] \quad (48)$$
$$+ \int_0^1 \frac{\tau \sin(t-\theta)}{F_r W_1} dz + W_1 \frac{\eta_{0\theta\theta\theta}}{r^3} \int_0^1 \int_1^{\bar z} \int_0^z \frac{W^2(z_2)}{W^2(z_1)W^2(z)} dz_2 dz_1 dz$$
$$+ \frac{2\tau W}{r}\sin(t-\theta)\int_0^1 \int_z^1 \left(\frac{U - (WI_2)'\eta_0}{W_1}\right) dz_1 dz$$
$$+ 2\tau \cos(t-\theta)\int_0^1 \int_1^z \left(\frac{(WI_2)'}{W_1}\frac{\eta_{0\theta}}{r}\right) dz_1 dz - 2\tau \cos(t-\theta)\frac{\eta_{0\theta}}{r} = 0.$$

The vertical integrals will be solved numerically using the trapezoidal rule. The final equation can be written in its dimensional form as:

$$A\eta_t + B\frac{\eta \eta_\theta}{r} + C\frac{\eta_{\theta\theta\theta}}{r^3} + D\frac{\eta_\theta}{r} + E\frac{\eta}{r} + F = 0, \quad (49)$$

where:

$$A = \left[\frac{2 + \int_0^{\bar{h}} \frac{W'}{W} \frac{dz}{H}}{\varepsilon^2 \Omega H}\right], \tag{50}$$

$$B = \frac{R}{\varepsilon^2 H^2}\left[-\frac{1}{W_1} - W_1' - \int_0^{\bar{h}}\left\{\frac{W'^2}{W} + \frac{2W'}{W^2} + \frac{1}{W^3}\right\}\frac{dz}{H}\right], \tag{51}$$

$$C = \frac{W_1 R^3}{\varepsilon H}\int\int\int_0^{\bar{h}\bar{h}\bar{h}+a}\frac{W^2(z_2)}{W^2(z)W^2(z_1)}dz_2 dz_1 dz, \tag{52}$$

$$D = \frac{2\tau W_1 \cos(\Omega t - \theta)}{\varepsilon \delta^2}\left[\int\int_0^{\bar{h}\bar{h}}\left(\frac{(WI_2)'}{W_1^2}\right)\frac{dz_1}{H}\frac{dz}{H} - 1\right], \tag{53}$$

$$E = \frac{2\tau W_1}{\varepsilon \delta^2}\sin(\Omega t - \theta)\int\int_0^{\bar{h}\bar{h}}\left(\frac{(WI_2)'}{W_1^2}\right)\frac{dz_1}{H}\frac{dz}{H}, \tag{54}$$

$$F = \int_0^{\bar{h}}\frac{\tau R \sin(\Omega t - \theta)}{HW_1 F_r}\frac{dz}{H} + \frac{2\tau R^2 W_1}{H}\sin(\Omega t - \theta)\int\int_0^{\bar{h}\bar{h}}\frac{U}{W_1^2}\frac{dz_1}{H}\frac{dz}{H}, \tag{55}$$

## 5 Numerical solution

KdV equation appeared so far in the analysis consists of nonlinear simple quadratic term, and dispersion term of the third order, that is evolving with time. For linear PDEs Fourier analysis is often used to obtain solutions or perform theoretical analysis. This is because the functions: $e^{i\xi x} = \cos(\xi x) + i\sin(\xi x)$ are essentially eigenfunctions of the differentiation operator $\partial_x = \frac{\partial}{\partial x}$. Differentiation of this function gives a scalar multiple of the function, and hence simple differential equations are simplified and can be reduced to algebraic equations as stated in [18]. The base function in case of Eq. (49) for example, is the free surface $\eta(\theta,t)$. Thus, we can write Fourier transform:

$$\eta(\theta,t) = \frac{1}{\sqrt{2\pi}}\int_{-\infty}^{\infty}\hat{\eta}(k,t)e^{ik\theta}dk. \tag{56}$$

The domain in which the motion is taking place is the outer circumference of the cylinder which forms the circle of 360°, thus the domain between: [0, 2π], with periodic boundary conditions, will be considered, thus the surface function satisfies the periodicity condition: $\eta(\theta + 2\pi) = \eta(\theta)$. Let us apply Eq. (56) into Eq. (49) we find for instance:

$$\eta_\theta = \frac{ik}{\sqrt{2\pi}}\int_{-\infty}^{\infty}\hat{\eta}(k,t)e^{ik\theta}dk, \tag{57}$$

$$\eta_{\theta\theta\theta} = \frac{-ik^3}{\sqrt{2\pi}}\int_{-\infty}^{\infty}\hat{\eta}(k,t)e^{ik\theta}dk. \tag{58}$$

By substituting into Eq. (49) we get:

$$\left[\frac{2 + \int_0^{\bar{h}}\frac{W'}{W}\frac{dz}{H}}{\varepsilon^2 \Omega H}\right]\eta_t = -\frac{ik}{2\pi r}\int_{-\infty}^{\infty}\hat{\eta}(k,t)e^{ik\theta}dk \tag{59}$$

$$* \int_{-\infty}^{\infty}\hat{\eta}(k,t)e^{ik\theta}dk \frac{R}{\varepsilon^2 H^2}\left[-\frac{1}{W_1} - W_1' - \int_0^{\bar{h}}\left\{\frac{W'^2}{W} + \frac{2W'}{W^2} + \frac{1}{W^3}\right\}\frac{dz}{H}\right]$$

$$-\frac{-ik^3}{\sqrt{2\pi}}\int_{-\infty}^{\infty}\hat{\eta}(k,t)e^{ik\theta}dk * \frac{W_1 R^3}{\varepsilon H}\int\int\int_0^{\bar{h}\bar{h}\bar{h}+\eta}\frac{W^2(z_2)}{W^2(z)W^2(z_1)}dz_2 dz_1 dz$$

$$-\frac{ik}{\sqrt{2\pi}}\int_{-\infty}^{\infty}\hat{\eta}(k,t)e^{ik\theta}dk * \frac{2\tau W_1 \cos(\Omega t - \theta)}{\varepsilon \delta^2}\left[\int\int_0^{\bar{h}\bar{h}}\frac{(WI_2)'}{W_1^2}\frac{dz_1}{H}\frac{dz}{H} - 1\right]$$

$$+\frac{1}{\sqrt{2\pi}}\int_{-\infty}^{\infty}\hat{\eta}(k,t)e^{ik\theta}dk * \frac{2\tau W_1}{\varepsilon \delta^2}\sin(\Omega t - \theta)\int\int_0^{\bar{h}\bar{h}}\frac{(WI_2)'}{W_1^2}\frac{dz_1}{H}\frac{dz}{H}$$

$$-\int_0^{\bar{h}}\frac{\tau R \sin(\Omega t - \theta)}{HW_1 F_r}\frac{dz}{H} - \frac{2\tau R^2 W_1}{H}\sin(\Omega t - \theta)\int\int_0^{\bar{h}\bar{h}}\frac{U}{W_1^2}\frac{dz_1}{H}\frac{dz}{H}.$$

The simple finite difference methods use forward Euler equation to discretize the time, which is one-step method, this means that all $\eta^{n+1}$ are determined from $\eta^n$ alone, we are going to divert from this case and use multistep method that involves other previous values. LeVeque [18] stated that the *leapfrog method* is midpoint method that mainly comes from Taylor expansion, for instance using the approximation:

$$\frac{\eta(t+dt) - \eta(t-dt)}{2dt} = \eta' + \frac{1}{6}dt^2\eta'''(t) + O(dt^3), \tag{60}$$

which can be written as:

$$\eta^{q+1} = \eta^{q-1} + 2dt f(\eta^q), \tag{61}$$

where the subscripts $q$ indicates to the time order. Which is a second order accurate explicit 2-step method, and it was also proposed in [19], they stated that the method is accurate for low enough wave numbers, but it loses accuracy rapidly for increasing wave numbers. At the leading order of the problem, we assume the other condition and discretize it in terms of the backward Euler difference to get rid of the values at the previous time:

$\eta = \bar{h} = const$ at $t = 0$, thus:

$$\frac{\partial \eta}{\partial t} = \frac{\eta_i^n - \eta_i^{n-1}}{\Delta t} = 0. \tag{62}$$

This leads to elimination of the value $\eta_i^{n-1}$ as:

$$\eta_i^{n-1} = \eta_i^n. \tag{63}$$

Concerning the integrations in the $z$-direction as mentioned are going to be integrated using the trapezoidal rule, one can find the information given in [20] for instance, where the integration is going to be written as:

$$\int_a^b f(z)dz \approx \sum_{i=0}^{n-1} W_i f(z_i), \qquad (64)$$

where $W_i$ are weights and $z_i$ the elevation points. The Trapezoidal method has the points:

$$z_i = a + ih, \quad h = \frac{b-a}{n-1}, \quad i = 0,\ldots,n-1. \qquad (65)$$

And the weights:

$$W_0 = W_{n-1} = \frac{h}{2}, \quad W_i = h, \quad i = 0,\ldots,n-2. \qquad (66)$$

Now the core function in Fourier transform is going to be converted into discrete Fourier space as used in [19]:

$$\hat{\eta}(k,t) = \frac{1}{N}\sum_{j=0}^{N-1} \eta(\theta_j,t)e^{-ik\theta_j}, \qquad (67)$$

$$-\frac{N}{2} \le k \le \frac{N}{2}-1. \qquad (68)$$

And then to get the inversion formula:

$$\eta(\theta_j,t) = \sum_{j=0}^{N-1} \hat{\eta}(k,t)e^{-ik\theta_j}, \quad 0 \le j \le N-1. \qquad (69)$$

Studying the stability using the leapfrog with Fourier discretization can be found in many references like in [19, 21]; the condition for the present KdV is deduced by taking the linearized version of KdV equation as follows:

$$\eta_t + \left(\frac{B+D}{A}\right)\frac{\eta_\theta}{r} + \left(\frac{C}{A}\right)\frac{\eta_{\theta\theta\theta}}{r^3} = 0, \qquad (70)$$

where A, B, C, D, are the coefficients introduced in Eq. (49). By assuming the base function of the form:

$$\eta(\theta,t) = f^{\frac{t}{\Delta t}} e^{ik\theta}. \qquad (71)$$

And then substitute it into Eq. (49) where the domain under study is already about the outer circumference, and the maximum wave number $k = \frac{\pi}{\Delta\theta}$, thus we get:

$$\frac{\Delta t}{(\Delta\theta)^3} < \frac{A}{C.\pi^3.r^3}. \qquad (72)$$

Now we are going to use data from the experiments where the wave was noticed. In the cases of relatively high volumes of water the noticed wave was smooth and symmetric in form but degenerated easily either flattened or broken, the cases here accord with volumes 12000 ml, and 14000 ml, and for tilts: $\tau = 0.0117$, $\tau = 0.0233$. However, in the cases of relatively small amounts of water the noticed wave was permanent single Kelvin wave that preserves its form for several rounds about the outer periphery, but we could not get the measurements of velocity with depth as the volume of water two small with the restriction of depth that are assumed for using the Vectrino. Fig. 4 and Fig. 5 show the real waves, it is clear that for relatively small water the wave crest appears little flattened, while increasing water amount the wave crest becomes more sharpened. The assumed scheme will take several cases including only big volumes, the initial value problem will be assumed based on the experiment using Gaussian function fit depending on the experimental data.

$$\eta_0 = A_m.e^{-\left(\frac{\theta-avg}{st}\right)^2}, \qquad (73)$$

where $avg$ the average of azimuthal data points, $st$ their standard deviation, $A_m$ the wave amplitude. We assume grid in the azimuthal direction consists of $m$ points, the ampli-tude has an angle about $\pi/2$, thus condition Eq. (72) is halved, for instance we present here case of volume 10000 ml, $m = 749$, $\Delta\theta = 0.001778$, $\tau = 0.01667$, $\Omega = 6.84$ rad/s, $\Delta t = 2.866 \times 10^{-06}$ s, the result in Fig. 6, where the other one in Fig. 7 is for the case of 12000 ml volume, $m = 899$, $\Delta\theta = 0.001594$, $\tau = 0.0117$, $\Omega = 6.02$ rad/s, $\Delta t = 6.1735 \times 10^{-06}$ s. Fig. 8 is for the case of 14000ml volume, $m = 649$, $\Delta\theta = 0.00224$, $\tau = 0.02333$, $\Omega = 6.82$ rad/s, $\Delta t = 1.753 \times 10^{-05}$ s. On tracking the wave with time, it turned out that the shear effect has only extinction effect on the wave amplitude that vanished gradually until it disappeared, however the shape of the wave was totally preserved to the starting guess, but never dispersed.

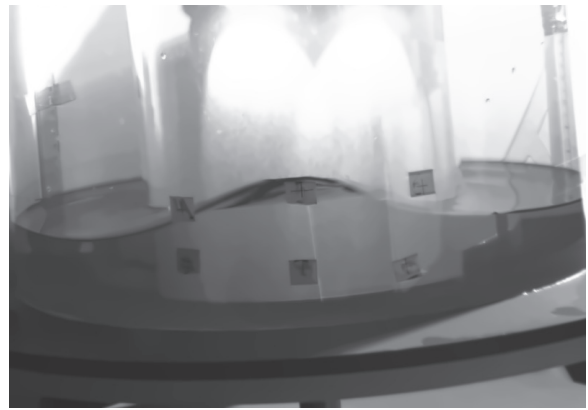

Fig. 4 Single Kelvin wave, $\bar{h} = 0.056$ m, $\tau = 0.0117$, $\Omega = 5.34$ rad/s

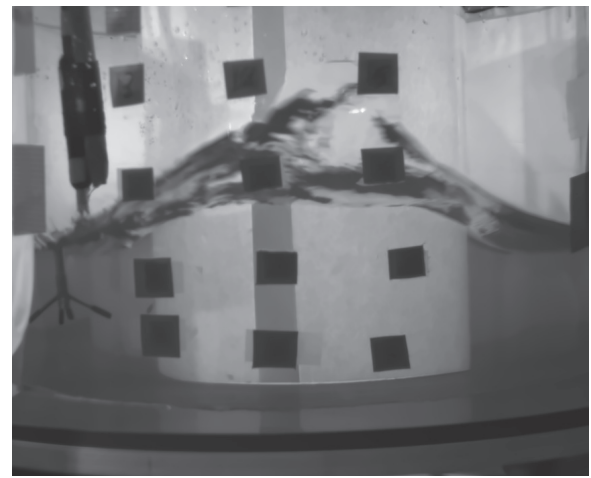

Fig. 5 Single Kelvin wave, $\bar{h} = 0.0933$ m, $\tau = 0.0167$, $\Omega = 6.84$ rad/s

The images went first under image processing steps starting from the calibration, and then pixel tracking, from which we could extract the exact pixels, which were connected with the real geometrical coordinates on the outer periphery of the cylinder, where there is created grid of small square paper clips in Fig. 4 and Fig. 5 those are benchmark points after interpolation and extrapolation techniques the pixels are fitted in between the real points and the real coordinates were extracted. For details about the image processing and other information can be found in [15].

This result is similar to the one extracted by the present author in other work still under revise taking the potential effect where it turned out that the shear effect has no implication on the shape of solitary kelvin wave moving in a frame of reference at the outer radius of the cylinder, Teles Da Silva and Peregrine [22] stated that to the solitary wave of high amplitudes and large Froude numbers encloses large regions of closed circulations and their shape appears to be insensitive to the vorticity distribution. In addition to this Brooke Benjamin [12] also mentioned that the vorticity in the stream has little effect on the wave in many circumstances typical of real open-channel flows, so that the results according to the potential theory which represents the stream as having a uniform velocity equal to the mean of the actual distribution, will often apply with good accuracy.

## 6 Conclusions

In this paper the shear effect on the flow in an open cylindrical channel under precession is discussed thoroughly. The treatment theoretically discussed the linear and the nonlinear parts of the problem, the linear case led to what we called the forced Rayleigh stability equation, for the inviscid version and to the forced Orr-Sommerfeld for the viscous one, as we use water in the experiment Rayleigh equation was discussed based on the stream function observation, as Coriolis force appeared in the equation we solved the equation at constant specific time and distance, which finally led to simple quadratic equation that had two different roots either complex or real, in any case the sign of those in the exponential part determines whether any disturbance will increase with time or simply vanish. The linear part of the problem also discussed other case after scaling the equations where new Burns' condition is derived, this connects between the stream velocity with the solid-body rotation one. Experimentally it turned out that the shape of the stream velocity distribution with depth is sinusoidal one, based on the control parameters of the problem which are the water volume in addition to the tilt angle and the rotation rate. By including this into the new forced Burns condition, the solution of any disturbance velocity has two different values relative to the bottom, similar to the normal conditions. Experimentally we could not access all volumes, because using the Vectrino needed effective depth to get good echo, but for the solitary wave only big amounts of water was needed, thus it was in our favor. On carrying on the derivation of the nonlinear part a new KdV model is derived, where the coefficients are all in terms of the shear function of the zeroth order. It was a question whether this shear affected the wave form or not, it turned out that the shear effect is limited and similar results to the extracted by the author in other work under irrotational conditions were extracted, which pour in favor other work like in [12, 22]. The solution for the new model of KdV depended on different discretizations, like in time it was the leapfrog method, and in distance the Fourier transform methods. Inserting the shear led to many integral with depth that were integrated using the simple trapezoidal rule.

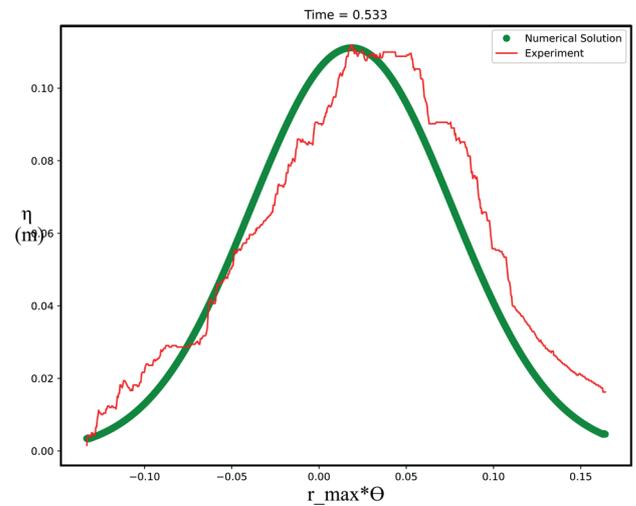

**Fig. 6** Single Kelvin wave taking the shear effect in the background, $\bar{h} = 0.0933$ m

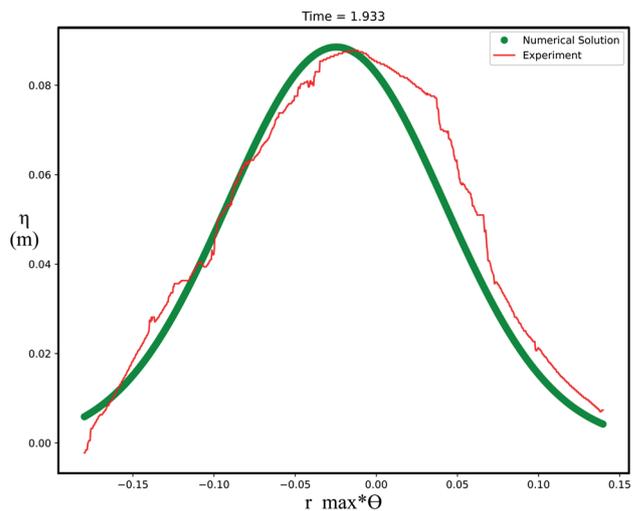

**Fig. 7** Single Kelvin wave taking the shear effect in the background, $\bar{h} = 0.0112$ m

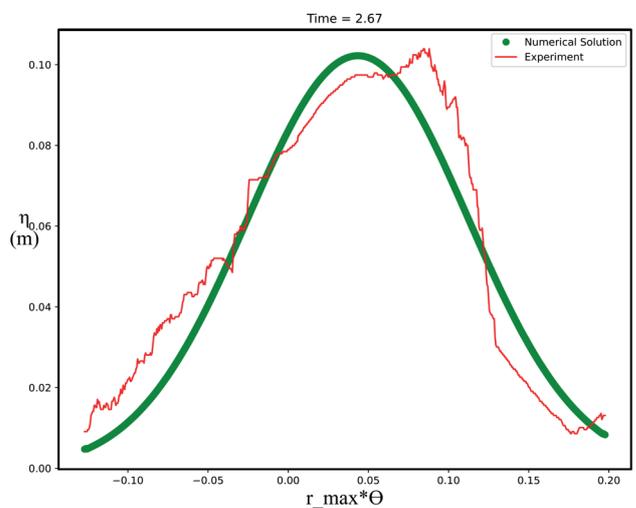

**Fig. 8** Single Kelvin wave taking the shear effect in the background, $\bar{h} = 0.1307$ m

**Acknowledgement**
This work is supported by the Stipendium Hungaricum Scholarship, under contract no. SHE-15651-001/2017. The author is thankful for the unknown referees for their invaluable comments on the early version of this manuscript.